\documentclass{aa}
\usepackage{graphics}
\usepackage{epsfig}

\def\beb{\begin{thebibliography}{999}}
\def\eeb{\end{thebibliography}}
\def\bei{\begin{itemize}}
\def\eei{\end{itemize}}
\def\bef{\begin{figure}}
\def\eef{\end{figure}}
\def\ben{\begin{enumerate}}
\def\een{\end{enumerate}}
\def\beq{\begin{equation}}
\def\eeq{\end{equation}}
\def\ber{\begin{eqnarray}}
\def\eer{\end{eqnarray}}

\begin{document}

\title{Unusual profile variations in pulsar PSR J1022+1001 \\ 
-- Evidence for magnetospheric ``return currents''?}

\author{R. Ramachandran \inst{1,2} \and M. Kramer \inst{3} }

\institute{Department of Astronomy, University of California at
            Berkeley, Berkeley, CA 94720, USA
\and Stichting ASTRON, PO Box 2, 7990 AA Dwingeloo, The
            Netherlands 
\and Jodrell Bank Observatory, University of Manchester, Macclesfield,
            Chesire SK11 9DL, UK}

\date{Received 31-03-03 / Accepted 01-07-03}

\abstract{We report a detailed multi-frequency study of significant
instabilities observed in the average pulse profile of the
16-millisecond pulsar PSR J1022+1001. These unusual profile variations
which are seen as a function of time and of radio frequency are
clearly different from classical profile mode-changing. We also note
discrete jumps in the polarisation position angle curve of this pulsar
which are remarkably coincident with the unstable profile
component. We propose that these jumps, as well as the instability of
the pulse profile, are due to magnetospheric return currents. This
would allow us to measure the basic properties of the magnetospheric
plasma for the very first time.
%
} 

\authorrunning{R. Ramachandran \& M. Kramer} \titlerunning{Profile
variations in PSR J1022+1001 -- evidence for return current?}

\maketitle

\section{Introduction}\label{intro}
\label{sec-intro}
The great success of pulsar timing and its numerous exciting
applications, e.g.~in the study of gravitational physics (e.g.~Taylor
1994), is often only possible due to the high precision which is
obtained during pulse time-of-arrival (TOA) measurements. The
procedure for obtaining TOAs involves cross-correlating the observed
time-tagged pulse profile with a high signal-to-noise standard pulse
profile ({\it template}) that is typically generated iteratively using
earlier observations. The implicit assumption is made that the average
pulse profile remains stable over a long time and small frequency
ranges. This assumption is usually well justified. In particular, a
few minutes of observation for millisecond pulsars (MSPs) leads to the
integration of several tens of thousand of single pulses, that is far
beyond the usual time scale of a few hundred to thousand pulses
necessary to obtain a stable pulse profile (e.g.~Helfand et al.~1975).
Until recently, only two fast rotating pulsars were known to exhibit
slow profile changes. These were PSR B1913+16 (Weisberg et al.~1989,
Kramer 1998) and PSR B1534+12 (Arzoumanian 1995, Stairs et al.~2000)
which both show profile changes on secular time scales due to
gravitational spin-orbit coupling.

This understanding that average profiles of MSPs are perfectly stable,
changed considerably with the observations of the 16-ms pulsar PSR
J1022+1001 (Camilo 1994) made with the Effelsberg radio telescope
(Kramer et al. 1999a, hereafter K99). This pulsar in a 7.8-d binary
orbit with a heavy CO white dwarf exhibits a characteristic
double-peaked average profile. The reported pulse shape changes are
most easily recognized by an alteration of the ratio of the two
component amplitudes.  This, of course, results in large TOA
variations when the pulsar is timed with a single standard template.
Using an {\it adaptive} template however, K99 were able to obtain a
timing accuracy that is comparable to other, ordinary millisecond
pulsars.  The detailed analysis by K99 furthermore showed that, out of
the two prominent components, only the leading, weakly polarised
component seems to undergo changes in intensity and structure, while
the trailing, highly polarised component remained stable, and arrived
``on time''.

While this different behaviour of the two components points to an
effect intrinsic to the pulsar magnetosphere, K99 also looked for
possible relationships of the profile changes to other effects, such
as time of day, parallactic angle, or even seasonal changes. None of
these, however, shows any correlation with the profile changes, and in
fact, a typical timescale for the pulse shape variations could not be
identified. An explanation for this appears to come from a totally
unexpected result obtained by K99.

Observations by K99 revealed that PSR J1022+1001 does not only show
profile variations as a function of time, but also on a narrow range
of radio frequency!  As K99 measured, the average profile at 1400 MHz
changes its shape on a typical frequency scale of about 8
MHz. Moreover, the modulation of the leading component resembles
closely to what one expect from interstellar scintillation, where the
bright ``scintils'' correspond to the flaring up of the leading
component. While this variation of the average profile as a function
of narrow frequency range points again to processes in the pulsar
magnetosphere, it indeed explains why no typical time scales for the
profile changes have been discovered yet; depending on the temporal
intensity maximum within the observing band, determined by the
ordinary interstellar scintillation, and the actual profile-frequency
pattern, different pulse profiles can be observed with different
timescales.

\begin{table}
\caption{List of observations made. The five columns give the date of
observation, centre frequency, bandwidth, number of frequency channels
and the effective sampling interval, respectively.}
\label{tab:obs_summ}
\begin{center}
\begin{tabular}{ccccc}\hline
{\bf Date} & {\bf $\nu_{\rm centre}$} & {\bf $\Delta\nu$} & {\bf N$_{\rm
chn}$} & {\bf $\Delta t_{\rm eff}$} \\ 
(dd:mm:yy) & (MHz) & (MHz) &  & ($\mu$s) \\ \hline\hline 
18-03-01 & 1380 & 80 & 512 & 51.4 \\
08-03-01 & 840 & 80 & 512  & 56.0  \\ 
08-03-01 & 328 & 10 & 512  & 113  \\ 
08-03-01 & 382 & 10 & 256  & 119  \\ 
22-03-01 & 1380 & 80 & 512 & 51.4  \\ 
22-03-01 & 328 & 10 & 512  & 113  \\ 
22-03-01 & 382 & 10 & 256  & 119  \\ 
22-03-01 & 840 & 80 & 512  & 56.0  \\ 
05-04-01 & 840 & 80 & 512  & 56.0  \\ 
05-04-01 & 1380 & 80 & 512 & 51.4  \\
05-04-01 & 328 & 10 & 512  & 113  \\ 
05-04-01 & 382 & 10 & 256  & 119  \\ \hline
\end{tabular}
\end{center}
\end{table}

Meanwhile, profile variations have also been reported for the
millisecond pulsars PSR B1821$-$24 (Backer \& Sallmen 1997), and PSR
J1730$-$2304 (K99), while significant changes in the polarisation
states have been identified for PSR J2145$-$0750 (Xilouris et
al.~1998, Sallmen 1998, Stairs et al. 1999). None of these sources has
been studied carefully for profile changes over a narrow frequency
range, since peak flux densities are usually smaller than for PSR
J1022+1001.

In this work we concentrate on PSR J1022+1001 in order to study the
extraordinary narrow frequency band profile variation in more
detail. We report observations of this pulsar made with the Westerbork
Synthesis Radio Telescope (WSRT) at four different centre frequencies,
i.e.~at 1380, 840, 382 and 328 MHz. These observations were motivated
by the work of Lyutikov \& Parikh (2000) and Lyutikov (2001) who
picked up the comments made by K99 and suggested that the profile
variations could indeed be caused by scattering of emission near the
light cylinder. In this case, it is obviously important to study the
narrow-band profile changes at various radio frequencies.

\begin{figure}
\begin{center}
\epsfig{file=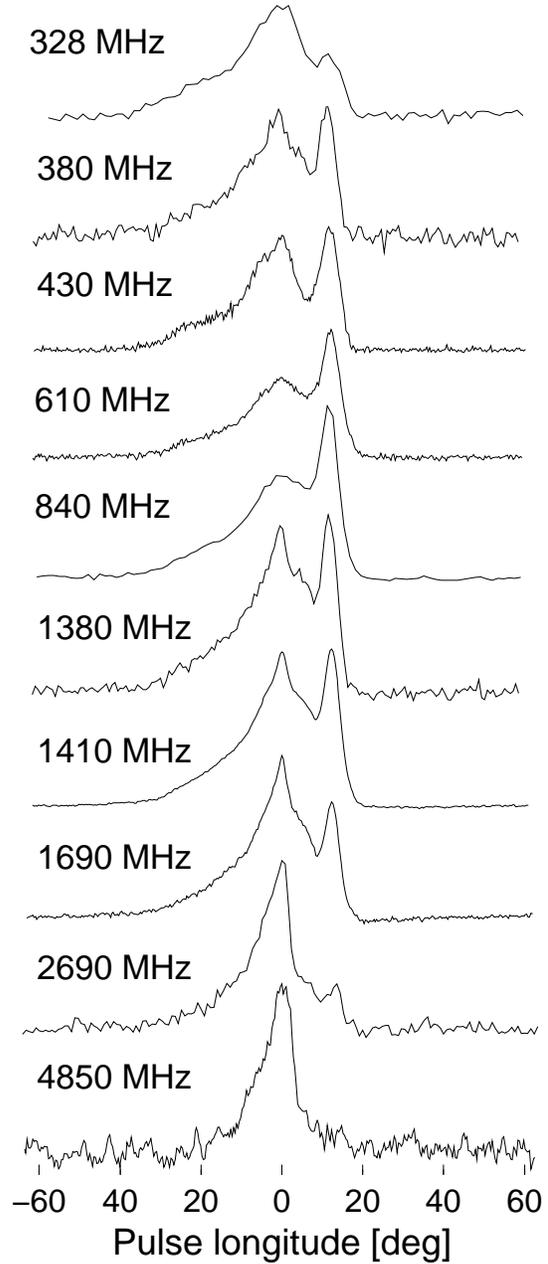,height=17cm}
\caption[]{Long term average profiles of J1022+1001 at various radio
frequencies. Profiles at 328, 382, 840 and 1380 MHz are from our
present observations. All the other profiles have been taken from
Kramer et al. (1999b) and references therein. Bandwidths of these
observations are (from 328 MHz to 4850 MHz) 10, 10, 10, 10, 80, 80,
40, 40, 80 and 112 MHz, respectively.}
\label{fig:avgprof}
\end{center}
\end{figure}

We also propose in this paper that the observed polarisation
properties of this pulsar are closely related to the instabilities
that we see in one of the components. As we will demonstrate, there is
a striking similarity between the observed average polarisation
position angle as a function of pulse longitude and the predictions of
Hibschman \& Arons (2001), who discuss the possible effect of
aberration and polar cap current flows on the observed polarisation
position angle curve. If this is true, then our results become the
first ever evidence for return currents in the pulsar magnetosphere.

After describing our observing set-up and procedures (Sect.
\ref{sec-obs}), we describe the observed characteristics of pulse
profile variation (Sect. \ref{sec-aveprof} --
\ref{sec-instability}). This includes discussion on their long term and
short term behaviour, and their possible relation to classical
``mode-changing''. In Sect. \ref{sec-polprop}, we describe the
observed polarisation properties of this pulsar and possible
explanations. This is followed by discussion and conclusions from this
investiation.


\section{Observations}
\label{sec-obs}
Observations were made in March to April 2001 with the pulsar backend,
{\tt PuMa}, at the WSRT.  The WSRT is an east-west array, with
fourteen equatorially mounted 25-m dishes. We have chosen this
telescope for its frequency agility combined with a powerful
state-of-the-art pulsar data acquisition system, and in particular we
have chosen it for its equatorial mount.

Even if only Stokes-I (total intensity) is recorded, there is in
principle a chance that severe polarisation cross-coupling effects
between signals from the two dipoles can affect the strength and shape
of the pulse profile. For instance, cross coupling can reduce the
observed total power of a randomly polarised signal by a certain
fraction, independent of the absolute orientation of the dipoles to
the plane of the sky.  For a completely linearly polarised signal, the
observed signal strength can vary as a function of the absolute dipole
orientation with respect to the direction of linear polarisation of
the signal.  When a partially linearly polarised source drifts in the
plane of the sky during the observation, the direction of linear
polarisation of the source signal incident on the antenna feeds
changes for alt-azimuth mounted telescopes.  For a telescope system
with severe cross-coupling errors, this may introduce some
time-dependence of the observed profile structure, as a function of
parallactic angle. As K99 demonstrate, this dependence is not observed
in the previous Effelsberg observations of PSR J1022+1001, but cannot
be fully excluded for the Arecibo observations also studied by
K99. Therefore, in order to exclude any such instrumental effect, we
choose to observe this pulsar with the WSRT.  With the WSRT being an
equatorially mounted telescope, the position angle of a linearly
polarised astronomical source remains constant with respect to the
orientation of the dipole as the source moves in the plane of the sky
during the observations. Any aforementioned effect with the potential
to alter the observed shape of the profile can therefore be excluded.

We used the WSRT at four different frequencies as detailed in
Table~\ref{tab:obs_summ}. For each observation, the delays between the
dishes were compensated and the signals were added ``in phase'' to
construct an equivalent 94-m single dish having an aperture
sensitivity of about 1.2 K Jy$^{-1}$. The frontends in all used bands
consisted of two orthogonal linear dipoles. In the WSRT signal
pipeline, the maximum allowed bandwidth at any given band per
polarisation channel is 80 MHz. In the intermediate frequency signal
path, this 80 MHz is divided into eight 10 MHz bands (whose individual
centre frequencies can be configured independently within the allowed
frequency range of the front end). For our 1380 MHz and 840 MHz
observations, we chose a contiguous frequency band of 1340--1420 MHz 
and 800--880 MHz, respectively.

Signals from each of the 10 MHz bands were Nyquist-sampled and Fourier
transformed to synthesize a filterbank of some specified number of
spectral channels. Finally, after some averaging, the 2-bit (4 level)
power samples were recorded in all the frequency channels. The final
effective sampling interval in each observation is given in Table
\ref{tab:obs_summ}. 

We compensated for the interstellar dispersion in each of these
observed data sequence by assuming a dispersion measure of 10.246 pc
cm$^{-3}$ (K99).

\begin{figure}
\begin{center}
\epsfig{file=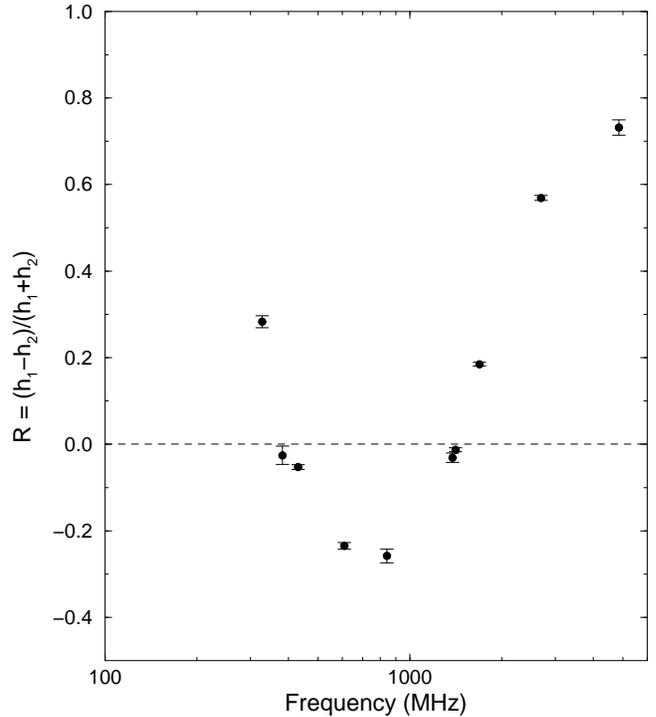,width=8.5cm}
\caption[]{Normalised component amplitude ratio $R$ as a function of
radio frequency.  A value of $0<R\le +1$ indicates a dominance of the
leading component, while for $-1\le R<0$ the trailing component is
stronger.  There is a strikingly systematic behaviour, with a minimum
at a frequency of about 800 MHz.}
\label{fig:ratio_freq}
\end{center}
\end{figure}


\section{Average profile evolution over large frequency range}
\label{sec-aveprof}
Average pulse profiles of pulsars are known to exhibit systematic
changes as a function of radio frequency. There are a number of
extensive studies in the literature which concentrate on this very
issue (e.g.~Rankin 1983; Hankins \& Rickett 1986; Lyne \& Manchester
1988, Kramer et al.~1994; Moffett \& Hankins 1996; Mitra \& Rankin
2002). A typical profile evolution would exhibit a narrowing of the
pulse width at higher frequencies, while at the same time the outer
profile components become stronger relative to the central
components. The most common understanding of frequency-dependent
systematic change is on the basis of the so called {\it
radius-to-frequency} mapping where radio emission of higher
frequencies are emitted closer to the neutron star than that of lower
frequencies (e.g.~Ruderman \& Sutherland 1975).  Mitra \& Rankin
(2002) give a recent complete summary of our understanding of these
effects.

Kramer et al.~(1999b) showed that this picture of ``typical'' profile
evolution cannot be applied to millisecond pulsars. In fact, they
pointed out that in most cases, profile evolution with frequency
is minimal for millisecond pulsars and that the profile width
hardly ever changes. They argue that the small size of a millisecond
pulsar magnetosphere does not allow a large stratification in emission
height and that in this case, the emission virtually comes from a single
location in the magnetosphere.

One notable exception to this distinctive millisecond pulsar behaviour
is again PSR J1022+1001, whose magnetosphere is however considerably
larger due to its relatively long spin-period of 16 ms. Kramer et
al.~(1999b) presented a number of multi-frequency profiles for this
pulsar, showing a remarkable frequency evolution. We improve their
summary here with additional profiles obtained from our observations
which are aligned by eye (see Fig.~\ref{fig:avgprof}).

The profiles shown are grand averages over all available observing
time and bandwidth (80 MHz in our case), which finally averages out
the effects which we will discuss in more detail in the following
section. Nevertheless, the profile evolution of this set of the grand
averages of PSR J1022+1002 is quite exceptional, and not easy to
explain.

At 328 MHz, the leading component is stronger. However, both
components stay almost equal in strength at 380 MHz and 430 MHz, after
which the trailing component becomes stronger. At 840 MHz the height
of the leading component is only about 60\% of the trailing
component. Above this frequency, the relative strength of the leading
component increases, and at the highest frequency (4848 MHz), the
trailing component is almost non-existent though still detectable.
This unusual behaviour is summarised very nicely in
Fig.~\ref{fig:ratio_freq}, where we have plotted a function of the
component amplitudes $h_{\rm 1}$ (leading) and $h_{\rm 2}$ (trailing), defined as
$R\equiv(h_{\rm 1}-h_{\rm 2})/(h_{\rm 1}+h_{\rm 2})$. $R$ has values between $+1$ and $-1$
depending on the relative strength of the leading and trailing
component. The plot of $R$ looks strikingly systematic in its
behaviour. Curiously, there is a minimum at radio frequency of about
800 MHz. This strange behaviour of $R$ with frequency clearly shows
that the two components exhibit entirely different frequency scalings.
The profile is hardly evolving over the frequency interval from 600 to
800 MHz and at very high frequencies. The profile evolves fastest at
frequencies around 500 MHz, 1400 MHz and very low frequencies.

This frequency evolution is unusual when compared to both normal pulsars
(e.g.~Mitra \& Rankin 2002) and millisecond pulsars
(Kramer et al.~1999b). It is obviously impossible to describe
the flux density spectrum of both components by a simple power law
across all frequencies. These results confirm 
the unusual spectral properties of both components which are
superposed on the profile variations at a single frequency.

\begin{figure*}
\begin{center}
\epsfig{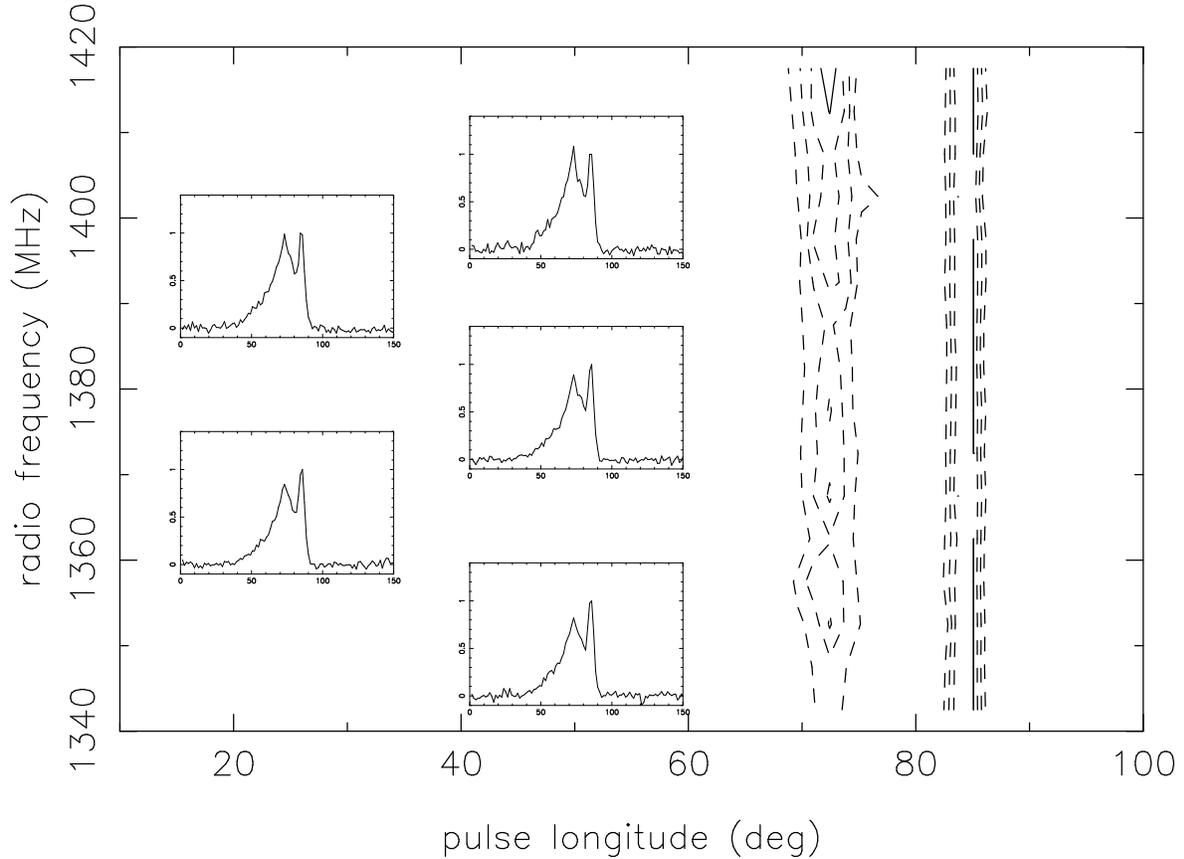}
\caption[]{Contour plot of pulse intensity as a function of pulse
longitude and observing frequency for an average pulse profile
constructed with 8 minutes of time series (approximately 30000 pulses)
in all the 16 channels each of 5 MHz width. For clarity, height of the
second component ($h_{\rm 2}$) has been normalised to unity in all the
average pulse profiles (right-side vertical contour strip). The five
inserts (from bottom to top) correspond to average profiles in channel
numbers 1, 5, 9, 13 and 16. Contour level steps are in terms of root
mean square error (see Eqn.~\ref{error}). `Solid' contour indicates
the region where $h_{\rm 1}/h_{\rm 2} \ge 1$. This figure has been plotted in the
same style as Fig.8 in K99 for easier comparison.}
\label{fig:long_freq}
\end{center}
\end{figure*}

\section{Profile Instabilities}
\label{sec-instability}
As K99 show, the leading component of the profile shows instabilities,
although the trailing component is very stable. In the following
sections, we study this in detail, both as a function of radio
frequency and time.

\subsection{Narrow-band variations}

In order to study the profile variations over a narrow frequency
range, we folded the de-dispersed time series in each frequency
channel at the period of the pulsar to generate average pulse
profiles.  An average pulse profile was produced for every specified
interval of time (typically 8 min) in each frequency channel. Each
time interval was long enough to have several tens of thousands of
rotation periods, so that we can expect that fluctuations of the
individual pulses as known for normal pulsars (Helfand et al.~1975)
are averaged out. 

Unfortunately, the pulsar was very weak during our observations at 328
MHz and 382 MHz. However, our observations at 840 MHz and 1380 MHz
were strong enough for us to investigate the phenomenon that we will
describe below.

Fig.~\ref{fig:long_freq} shows the contour plot using the 1380 MHz
data set of average pulse profiles as a function of radio
frequency. The two vertical trails of contours correspond to the two
components of the average pulse profile. This plot corresponds to a
time interval of 8 min ($\sim$ 30,000 individual pulses). Although the
original observation had 512 frequency channels across the 80 MHz
bandwidth, after de-dispersion, due to signal-to-noise ratio
considerations, we have added 32 adjacent channels, so that the band
is represented by only 16 channels. For clarity, the average pulse
profiles in all the frequency channels are normalised in such a way
that the height of the stable trailing component is unity (right-side
vertical contour strip). However, the leading component, as the
contours (as well as the five inserted plots) show, is unstable. The
five inserted average profiles (from bottom to top) correspond to
frequency channels 1, 5, 9, 13 and 16, respectively. Contour levels
have been plotted in steps of root mean square error
(Eqn.~\ref{error}). The observed profile variations are perfectly
consistent with the results presented by K99 using a different
telescope with different polarisation feeds, different hardware, and
in particular with a different telescope mount.

It is important to appreciate that this instability of the leading
component is not due to interstellar scintillation, for it affects
both the components equally. Since we have normalised the height of
the trailing component to unity, the relative variation seen in the
leading component must be independent of interstellar scintillation
effects and intrinsic to the pulsar magnetosphere.

\begin{figure*}
\begin{center}
\epsfig{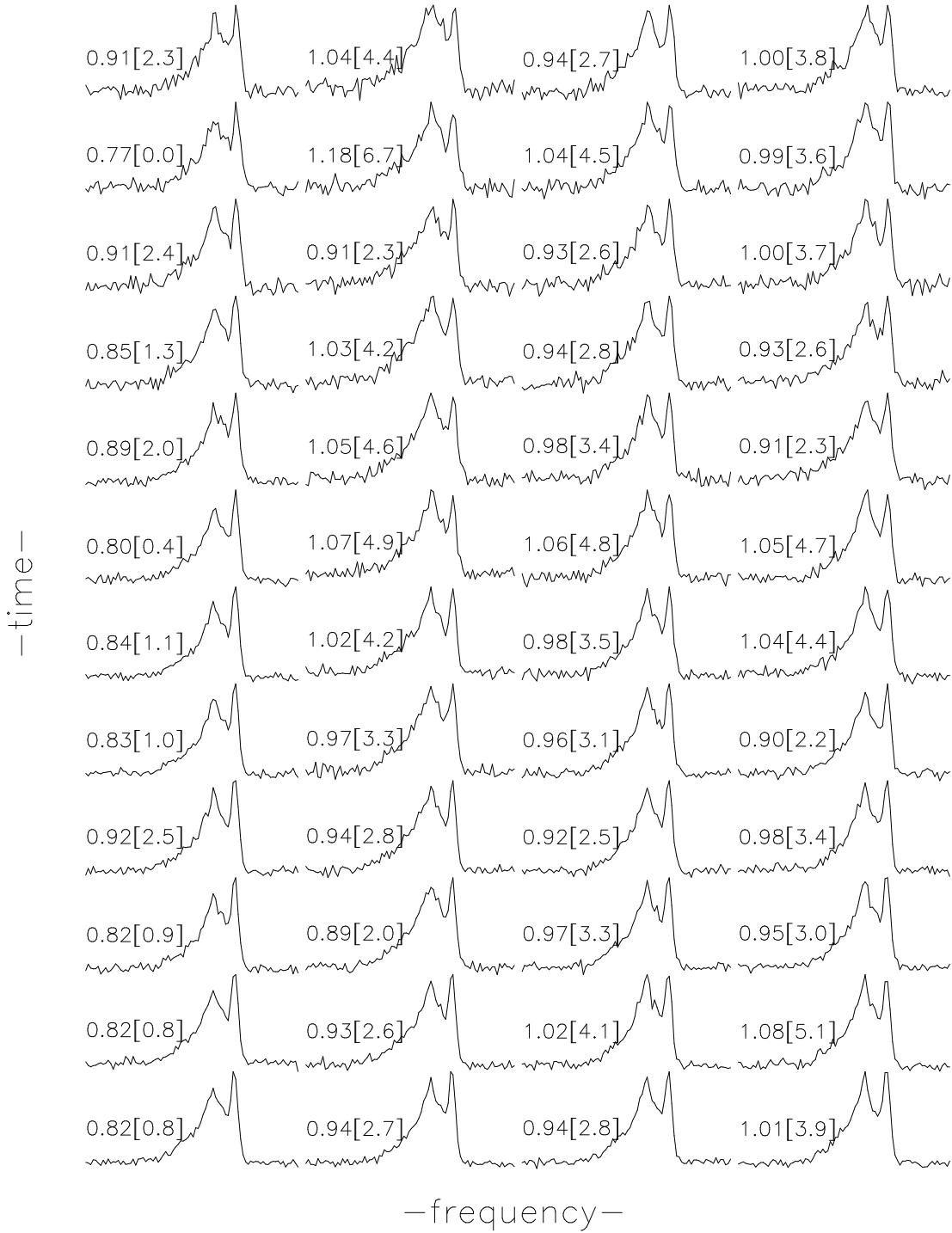}
\caption[]{Average pulse profiles as a function of time in four
different frequency channels. Each average profile was created with 8
minutes of time series. The four columns correspond to frequency
channels with centre frequencies of 1382.5, 1407.5, 1412.5 and 1417.5
MHz, respectively, with a channel width of 5 MHz. The fractional
height of the first component ($h_{\rm 1}/h_{\rm 2}$) is also given for each
profile, along with its deviation from the lowest ratio value
(1$^{st}$ column, 2$^{nd}$ profile from top, ratio of 0.77) in terms
of root mean square error (Eqn.~\ref{error}). See text for details.}
\label{fig:timevar}
\end{center}
\end{figure*}

\begin{figure*}[th]
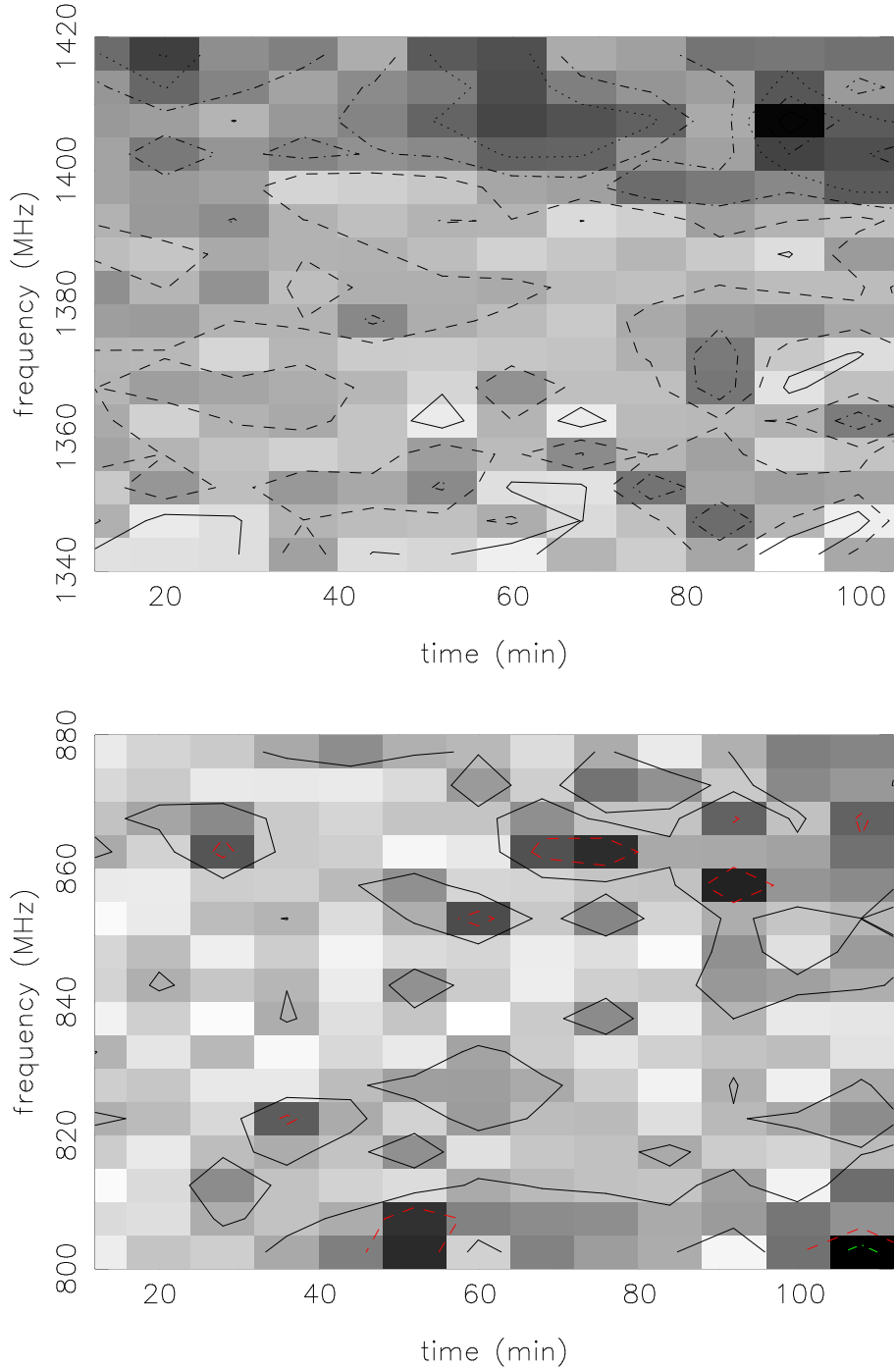

\begin{center}
\epsfig{file=fig5a.ps,height=12cm,angle=-90}

\vspace*{0.5cm}

\epsfig{file=fig5b.ps,height=12cm,angle=-90}
\caption[]{Component amplitude ratio as a function of time (X-axis)
and radio frequency (Y-axis). Each point corresponds to a time length
of 8 min. Top panel corresponds to the observation at 1380 MHz, and
the bottom panel to 840 MHz. Contour levels, in steps of root mean
square error (see Eqn.~\ref{error}), are indicated with different line
styles for clarity. In the top panel, from the lowest contour level
(white colour), the line style goes as `solid', `dash', `dot-dash',
`dot-dot-dash', `dotted', and `solid'. The highest contour level
corresponds to the darkest gray scale point. In the bottom panel, as
the signal to noise ratio is not as high as that of the top panel,
there are only three significant contour levels.}
\label{fig:time_freq_gray}
\end{center}
\end{figure*}

\subsection{Instabilities as a function of time}
In order to study the profile changes as a function of time in more
detail, we divided, as described above, our total observation time
length into small segments of approximately 8 minutes in length, in
each of the 16 frequency channels. With this set of profiles,
normalized again to the trailing component, variations were also seen
as a function of time. In Fig.~\ref{fig:timevar} we show average
profiles of these 8 minute segments stacked from bottom to top, in
four different frequency channels (each of 5 MHz width). We also show
the relative height of the first component with respect to the second
($h_{\rm 1}/h_{\rm 2}$), and the statistical significance of this value with
respect to the minimum value of ($h_{\rm 1}/h_{\rm 2}$) in the data set
($=\;0.77$) in square brackets. These significance values have been
calculated with the error in the amplitude ratio, $\Delta q$, which is
calculated as
\begin{equation}
\label{error}
\Delta q = \Delta h \;\sqrt{ 1 + q^2 } \approx 
\sqrt{2}\Delta h
\end{equation}
where $\Delta h$ is the error in component amplitude, taken to be one
off-pulse RMS. From Fig.~\ref{fig:timevar} it is clear that the
observed profile variations are statistically significant.

Having shown that the profile variations occur both in time and narrow
range of frequency, we now attempt to demonstrate the variation in
both dimensions of the parameters simultaneously.
Fig.~\ref{fig:time_freq_gray} shows the gray scale (and contour) plot
of this variation as a function of time (in X-axis) and radio
frequency (in Y-axis). With a ``pixel'' size of 8 min$\times$5 MHz, we
again measure the ratio of the amplitudes of the leading and trailing
components in each of the average profiles relative to its long-term
average, and plot this number in a gray scale plot.  The additionally
drawn contours are chosen in steps of root mean square error, as
computed again from Eqn.~\ref{error}. The solid and dashed line
contours indicate amplitudes of the leading component being larger or
smaller than the amplitude of the long-term average, respectively. The
two panels (top \& bottom) of Fig.\ref{fig:time_freq_gray} corresponds
to our 1380 MHz and 840 MHz observations. As we can see in the top
panel where the pulsar was bright, there are 6 contour levels,
indicating that the fluctuations are statistically significant.

\subsection{Time scales, frequency scales and the distribution 
of height ratios}
\begin{figure*}
\begin{center}
\epsfig{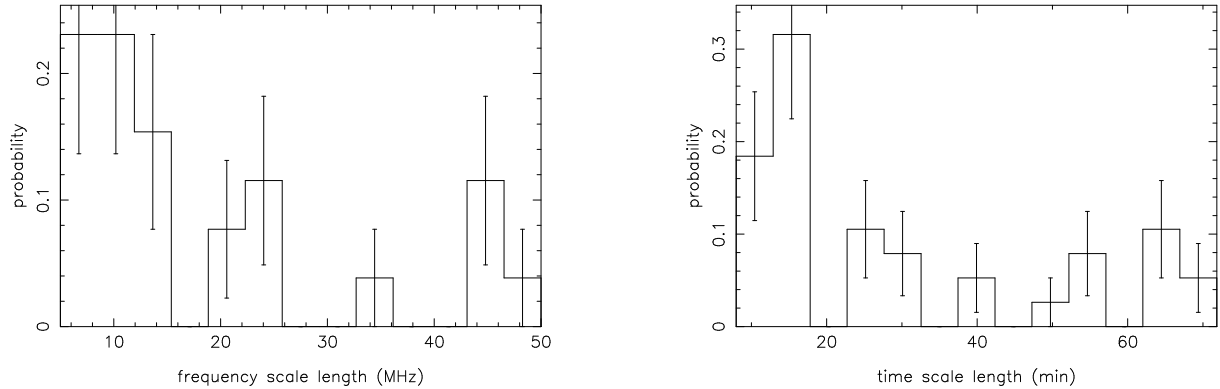}
\caption[]{Distribution of frequency and time scale lengths. Only the
data set at 1380 MHz was used for this plot. As the plots show, a
range of frequency and time scales are seen.}
\label{fig:f_t_scale}
\end{center}
\end{figure*}

Our previous analysis and that of K99 suggests the lack of a
systematic trend in the variations as a function of time or frequency.
In order to investigate in more detail as to whether there is a
preferred quasi-periodicity or any preferred time or frequency scale,
we computed histograms from observed changes in frequency and
time. The results are summarised in Fig.~\ref{fig:f_t_scale}. We computed 
these variations by measuring the half-maximum points of the ratio 
variations along frequency and time axes in Fig.~\ref{fig:time_freq_gray}.
That is, for the left panel in \ref{fig:f_t_scale}, frequency scale 
for each time pixel was measured, and for the right panel, time 
scale for each frequency channel was measured. It is clear that 
no clear tendency is observed. We also computed two-dimensional 
auto-correlation functions of the data shown in 
Fig.~\ref{fig:time_freq_gray} with  similar results.

Even though it is not possible to derive characteristic frequency or
time scale, it is important to emphasize that the observed profile
variations are highly significant. Both Fig. \ref{fig:timevar} \&
\ref{fig:time_freq_gray} show variations relative to high
signal-to-noise templates which can be detected with high
significance. Hence, the gray scale plots must not be mistaken for
``noisy maps''. The appearance is merely a different representation of
the fact that typical time or frequency scales cannot be identified.

\begin{figure*}
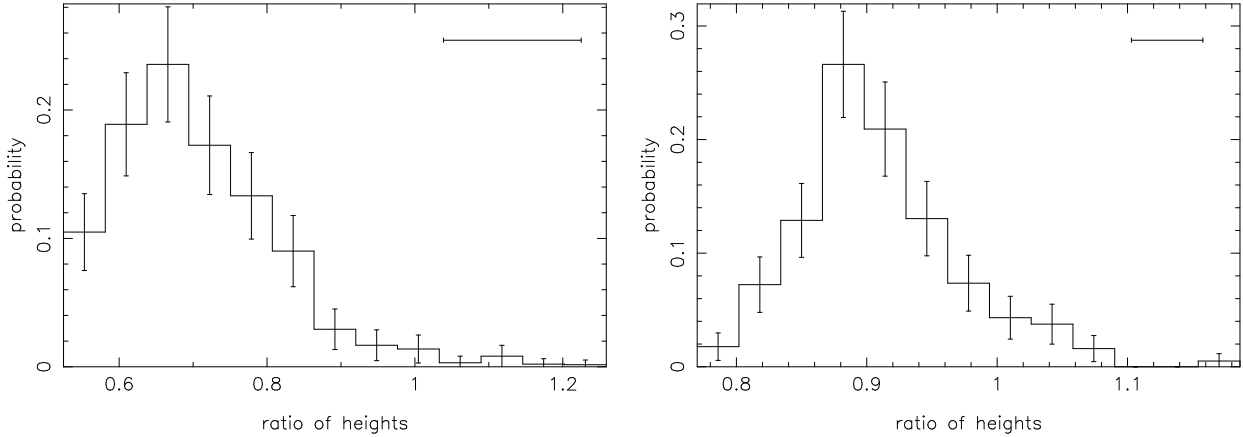

\begin{center}
\begin{tabular}{cc}
\epsfig{file=fig7a.ps,height=8cm,angle=-90}
&
\epsfig{file=fig7b.ps,height=8cm,angle=-90}
\end{tabular}
\caption[]{Distribution of height ratios of the two components
observed at 840 MHz (left) and 1380 MHz (right), for each 8
min. average profile in each 5 MHz frequency channel. Horizontal error
bar in the top right corner indicates 1$\sigma$ error.
The total extent of the X-axis (from one extreme height
ratio to the other) spans over several times the error bar, and hence
the profile variations are statistically significant. See text for
details.}
\label{fig:distrib_dh}
\end{center}
\end{figure*}

Having discussed the time scales and frequency scales, we also give in
Fig.~\ref{fig:distrib_dh} the distribution of component height ratios.
These component ratios were computed from the short-term 8 minute
average profiles in each of the 5 MHz frequency channels. One of the
possibilities for the observed profile variations is classical
``mode-changing'' (Backer 1970), which is well known among the normal
pulsar population. Usually, if the mode-changing time scale is much 
shorter than the averaging time, averaging of several thousands of
pulses is enough to get a stable profile even with the classical mode
change. In the case of PSR J1022+1001, even after taking 8 minutes
average (30,000 pulses), we still see variation. In order for
mode-changing to have an effect on profile variation, then the time scale
of these modes needs to be of the order of several minutes (less 
than 8 minutes). 

As described by Kramer et al. (1999a), timing measurements obtained
with the help of ``adaptive'' template (by gaussian component fitting
method) gives the best timing residuals, and indicates that the
variation that we see is much smoother than what one would expect from
a strictly bimodal distribution.  Nevertheless, from our data set
alone we cannot rule out a case of classical mode-changing, e.g.~where the
strength of the primary mode dominates over a weaker, less probable
secondary mode. However, the narrow-band profile variations as
a function of frequency observed here and by Kramer et al.~(1999a)
cannot be explained by mode-changing. 


\section{Polarisation properties}
\label{sec-polprop}
\begin{figure}
\begin{center}
\epsfig{file=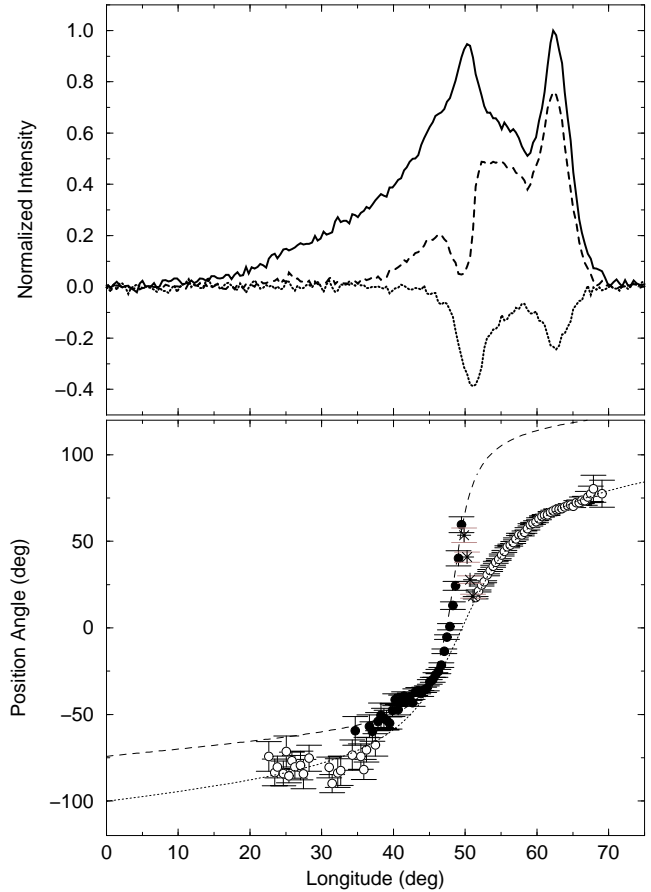,width=9cm}
\caption[]{Top panel gives the average pulse profile of PSR J1022+1001
at 1420 MHz (solid line), the average linearly polarised intensity
(dashed line), and the average circularly polarised intensity (dotted
line). The bottom panel shows the position angle of the linearly
polarised emission component as a function of pulse longitude. See
text for details.}
\label{fig:pafit}
\end{center}
\end{figure}

In addition to the unusual instability in the pulse profile, PSR
J1022+1001 has unusual polarisation properties. We have given in
Fig. \ref{fig:pafit} the observed average linear polarisation position
angle. This figure is a reproduction of Fig. 9 of K99 (refer to K99
for details on observing set-up.) In the top panel, we show the total
intensity as well as the linearly and circularly polarised
intensities. The PA curve is rather peculiar as it shows a step at
about 35 deg longitude and a ``notch'' at about 50 deg longitude.

We have split the PA curve at the positions of the step and the notch
into two separate parts, indicated by filled and open circles,
respectively. Most remarkably, both resulting parts of the PA curve
can be described by standard Radhakrishnan--Cooke (RC) models
(Radhakrishnan \& Cooke 1969) which only differ in their $\beta$
value, i.e.~angle of closest approach between the magnetic axis and
the line of sight. Indeed, with $|\beta|$s of $1.5^\circ$ and
$4.5^\circ$, respectively, both RC models have the same $\alpha$ value
(angle between the rotation and the magnetic axis) of $50^\circ$
although this value is not very well constrained.  The PA data
intermediate between the two PA curves, indicated by ``star'' symbols,
were ignored for the fit.  An RC-model has also been fitted to PSR
J1022+1001 by Xilouris et al.~(1998) and Stairs et al.~(1999), both
attempting to fit the whole PA curve, but essentially ignoring most of
the region designated by filled symbols in Fig.~\ref{fig:pafit}.
Consistent with our results, at 1410 MHz Xilouris et al.~(1998) derive
values of $\alpha=53.2^\circ$ and $|\beta|=7.3^\circ$, while Stairs et
al.~(1999) obtain at 610 MHz $|\beta|=4.9^\circ\pm1.8^\circ$ and at
1410 MHz $|\beta|=7.1^\circ\pm0.3^{\circ}$.

One of the possible interpretations for this very interesting PA curve
is to view these two parts as being caused by two polarisation
emission modes (which may not be orthogonal) that arise from two
different altitudes in the magnetosphere. In a cylindrical geometry,
the inferred value of $\beta$ (and $\alpha$) for the two modes cannot
be different, even if they are emitted from two different
altitudes. However, as the emission altitudes of radio emission from
MSPs are likely to be a good fraction of the light cylinder radius
(Kramer et al. 1998), two effects may become important: aberration of
the pulsar beam in a forward direction (i.e., earlier arrival times),
and bending of the emission beam due to a sweep-back of the magnetic
field lines near the light cylinder (i.e.,~later arrival
times). Typically, aberration is the more prominent effect, in
particular for fast rotating MSPs, unless the emission is originating
from close to the light cylinder (e.g.~Phillips 1992, Kramer et
al.~1997). As the net result, if the emission corresponding to the two
modes arises from two different altitudes, the derived values of
$\beta$ for the two modes could be different. The radiation associated
with each component would then propagate through different parts of
the pulsar magnetosphere.  Depending on relative height and the
different viewing geometries, described by the determined
$\beta$-values, the leading component could hence be subject to
(different) propagation effects which may then affect its apparent
stability.

The intermediate PA-values (``star'' symbols) could be explained by an
incoherent addition of the two modes which are {\em not} orthogonal,
as it has been observed for other pulsars before (e.g.~McKinnon
2003). Curiously, this happens at longitudes which exactly coincide
with the unstable leading component of the profile. We note that the
idea that polarisation modes may be emitted from different heights is
not new. McKinnon \& Stinebring (1998; 2000) showed that the average
pulse profiles corresponding to two orthogonal polarisation modes are
different in their widths, indeed suggesting that modes are emitted at
separate altitudes. This effect has been further explored and modelled
by Rankin \& Ramachandran (2003) with a large sample of pulsars.

\begin{figure}
\begin{center}
\epsfig{file=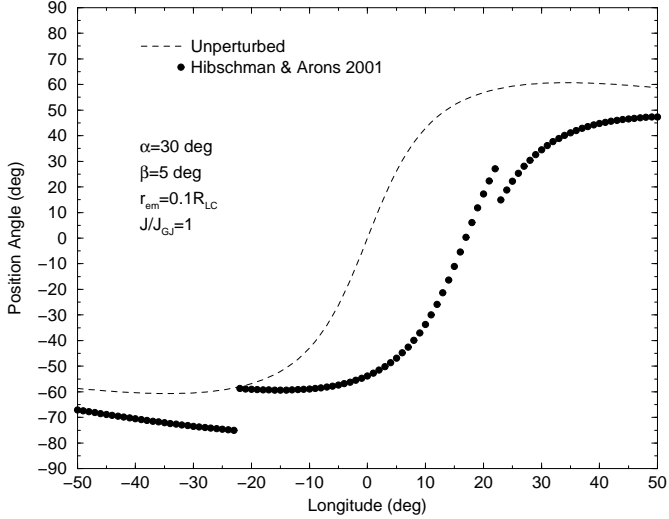,width=7cm,angle=-90}
\caption[]{Linear polarisation position angle as a function of pulse
longitude. `Dash' line shows the unperturbed curve, and the dotted
line shows the curve after taking into account the effect of
aberration and magnetospheric return currents. $r_{\rm em}$ \& $R_{\rm
LC}$ indicate emission height and light cylinder radius, respectively,
and $(J/J_{\rm GJ})$ is the current in terms of Goldreich-Julian
value. (reproduction of Fig.~5 of Hibschman \& Arons 2001). The point
with longitude value zero represents the closest approach of magnetic
pole to the line of sight.}
\label{fig:Hibschman}
\end{center}
\end{figure}

Another independent explanation for this peculiar PA behaviour, which
may be more plausible, is derived from the work by Hibschman \& Arons
(2001). Following ideas first developed by Blaskiewicz et al.~(1991),
they examine the effect of aberration on the RC model curve, but also
include polar cap current flows in their calculations.  As they
summarise, the current flow tends to shift the PA curve upwards (in
Fig. \ref{fig:pafit}), while aberration tends to shift the PA value in
the opposite direction, causing also a phase shift of the whole PA
curve. Most importantly, they also consider the possible effect of a
return current in the pulsar magnetosphere. They argue that if our
line-of-sight happens to penetrate a layer of this return current,
this would appear as ``jumps'' in the PA curves at the corresponding
pulse longitude values. In fact, Fig.~5 of Hibschman \& Arons 2001
(reproduced here in Fig. \ref{fig:Hibschman}) shows very remarkable
similarity to our measured PA curve shown in
Fig. \ref{fig:pafit}. Their theoretical curve computed for an assumed
set of parameters of $\alpha=30$ deg, $\beta=5$ deg and an emission
height of 10\% of the light cylinder radius, shows both a step as well
as a notch, similar to what is actually observed for PSR J1022+1001.

Given this striking similarity between this theoretical prediction and
our actual observed data, we are tempted to conclude that the discrete
jumps that we see in the position angle are not related to orthogonal
modes, but due to the return current layer as suggested by Hibschman
\& Arons (2001).  If this is true, this is the first ever
observational evidence for a return current in a pulsar magnetosphere.

The discrete jumps caused by the return current may be, as the authors
indicate, frequency-dependent, depending on the nature of
radius-to-frequency mapping, and the total height of the emission
region. Under these assumptions, in order derive definitive
conclusions about the nature of the return current layer, one needs to
study all the characteristics of the PA jumps, including their
frequency dependence. We are currently undertaking such a detailed
study, and the results will be presented in a following publication
(Kramer, Ramachandran, Stairs \& Athanasiadis, in prep.). Preliminary
fits of the Hibschman \& Arons model to our PA data suggest an
emission height of 40\% of the light cylinder radius and a current of
about 75\% of the Goldreich-Julian value, although further studies are
needed to confirm these values.

As one can see, the unstable leading component in the pulse profile
directly corresponds to the region bracketed by the discrete PA jumps
at longitudes $\sim35$ deg and $\sim50$ deg (Fig. \ref{fig:pafit}),
which, in the above model, mark the longitudes of the return current
layer. We propose here that the instability in the component strength
is therefore due to the illuminated return current layer in the pulsar
magnetosphere. Hibschman \& Arons (2001) point out that although the
return current layer is not normally thought of as a site of emission,
it may be visible either through direct emission or through refraction
or scattering of radiation of normal pulsar emission. While the
physical reason for the instability of the component itself is not
clear, the fact that the component appears at the same location as the
discrete jump of the PA curve, and the similarity to the theoretical
predictions of Hibschman \& Arons (2001) may not be just a
coincidence.

\section{Discussion and Conclusions}
\label{sec-discuss}
Our observations were conducted with the Westerbork Synthesis Radio
Telescope (WSRT). In all our observing bands, WSRT has a dual-linear
dipoles. The telescope is equatorially mounted, and hence the relative
orientation of the direction of linear polarisation of the source
remains the same, irrespective of the position of the source in the
sky. This helps us in eliminating time-dependent profile variations
due to possible residual cross-coupling effects. Moreover, as K99
showed (Fig.~9; see also Sallmen 1998 and Stairs et al.~1999), the
degree of polarisation of the leading component is significantly less
than that of the trailing component. If there is any spurious
instrumental effect related to polarisation cross-coupling, we would
expect more variations in the trailing component - in contrast to what
we observe.

Therefore, without doubts we have confirmed the profile variations
observed in PSR J1022+1001 as a function of time and, in particular,
radio frequency. These variations are highly significant, as we have
shown in Figs. \ref{fig:long_freq} \& \ref{fig:time_freq_gray}. As for
K99, this behaviour has been clearly observed in the 1340--1420 MHz
band and is, for the first time, also clearly detected in at least
another frequency band, i.e.~800--880 MHz. Due to signal-to-noise
constraints, it was not possible to confirm these variations also with
our 328 and 382-MHz frequency observations.

Even though Fig. \ref{fig:timevar} \& \ref{fig:time_freq_gray} are
reminiscent of an interstellar scintillation pattern, the relative
variation of component amplitudes eliminates the possibility of a
propagation effect outside the pulsar magnetosphere, as scintillation
must affect both the components equally.

The observed profile changes does not seem to be consistent with {\it
profile mode-change}, which is well known for the ordinary long period
pulsar population. A smoothly distributed component ratios
(Fig. \ref{fig:distrib_dh}) and their narrow band variations seem to
rule out this possibility.

Lyutikov \& Parikh (2000) and Lyutikov (2001) discuss the existence of
a scattering and diffracting screen close to the light cylinder of
fast rotating pulsars. While their estimates of the diffractive scales
can vary within order of magnitude, they note that if the scattering
is due to turbulence inside the pulsar magnetosphere it should be
independent of frequency. Due to the apparent range of frequency
scales suggested by Fig.~6 and 7, both at 840 and 1400 MHz, it is
difficult to test this prediction, although the profile variations
shown in Fig.~6 appear similar at the first glance. However, the
frequency scalings may be different for other types of scattering
(Lyutikov, private communication) and it seems difficult to either
confirm or reject the scattering screen model.

It is clear from Fig.~\ref{fig:ratio_freq} that
the spectral properties of the unstable leading component are very
different from that of the stable trailing component. Although the
reason for this striking dissimilarity is unclear, it may well be
related to the nature of the return current layer.

If our conclusions are correct, one would expect that this effect
should also be visible for other pulsars. However, it is important to
realize that for the involved aberration effect to become significant,
relatively large emission heights are required, i.e.~typically 10\% of
the light cylinder radius or more. For normal, slowly rotating
pulsars, estimates for emission altitudes are usually much lower
(e.g.~Blaskiewicz et al.~1991, Phillips 1992, Kramer et
al.~1997). However, for MSPs the emission appears to originate at
larger fractions of the light cylinder radius (Kramer et al.~1998), so
that these objects are indeed those where one would expect to observe
the described effects at first. Indeed, a large fraction of MSPs is
known to exhibit PA curves which are significantly different from what
is expected for a classic RC curve (Xilouris et al.~1998, Stairs et
al.~1999).  A detailed study of MSP polarimetry data is part of the
study currently underway (Kramer et al.~in prep.)

\section{Summary}
With new multi-frequency data, we have demonstrated that the
peculiar profile variations of PSR J1022+1001 over time and a narrow
range of frequency can be observed at widely spaced radio frequencies.
We propose two alternative models which are capable of explaining this
interesting phenomenon. If the model currently favoured by the data is
correct, our study would be the first ever observational evidence for
a return current in a pulsar magnetosphere. If confirmed, future
detailed studies will then allow measurements of basic properties of the
plasma flow in the magnetosphere for the very first time.

\acknowledgements We thank Jon Arons, D. C. Backer, A. G. Lyne,
G. Smith and I. Stairs for all the discussions and critical
comments. We also thank our referee, Simon Johnston, for his critical
comments that have helped improve the manuscript.

\end{document}